\documentclass[12pt]{article}
\newsavebox{\ns}
\newsavebox{\dbrane}
\newsavebox{\dbshort}

\usepackage{epsfig}
\usepackage{amssymb}

\def\appendix{{\newpage\section*{Appendix}}\let\appendix\section%
        {\setcounter{section}{0}
        \gdef\thesection{\Alph{section}}}\section}

\def\be{\begin{equation}}
\def\ee{\end{equation}}
\def\ba{\begin{eqnarray}}
\def\ea{\end{eqnarray}}

\def\Dslash{\,\,{\raise.15ex\hbox{/}\mkern-12mu D}}
\def\Dbarslash{\,\,{\raise.15ex\hbox{/}\mkern-12mu {\bar D}}}
\def\delslash{\,\,{\raise.15ex\hbox{/}\mkern-9mu \partial}}
\def\delbarslash{\,\,{\raise.15ex\hbox{/}\mkern-9mu {\bar\partial}}}
\def\pslash{\,\,{\raise.15ex\hbox{/}\mkern-9mu p}}
\def\calDslash{\,\,{\raise.15ex\hbox{/}\mkern-12mu {\cal D}}}

\newcommand\mr{\mathbb{R}}
\newcommand\mc{\mathbb{C}}
\newcommand\mz{\mathbb{Z}}
\newcommand\mcp{\bf \mathbb{C}P}
\newcommand\mhp{\bf \mathbb{H}P}

\newcommand\Index{\mathrm{Index}}

\newcommand\cosec{\mathrm{cosec}}

\newcommand\Pf{\mathrm{Pf}}
\newcommand\D{\mathcal{D}}
\newcommand\E{\mathcal{E}}
\newcommand\diff{\mathrm{d}}

\begin{document}

\begin{titlepage}

\begin{center}
\today
{\small\hfill hep-th/0310147}\\
{\small\hfill Imperial/TP/03-4/1}\\

\vskip 1.5 cm
{\large \bf Global Worldsheet Anomalies from M-Theory}

\vskip 1 cm
{James Sparks}\\
\vskip 1cm

{\sl The Blackett Laboratory \\
Imperial College London \\
Prince Consort Road \\
London SW7 2AZ, U.K. \\
\

{\tt j.sparks@imperial.ac.uk}\\}

\end{center}

\vskip 0.5 cm
\begin{abstract}

We present an M-theory proof of the anomaly of Freed and Witten
which in general shifts the quantisation law for the $U(1)$ gauge
field on a D6-brane. The derivation requires an understanding of
how fields on the D6-brane lift to M-theory, together with a
localisation formula which we prove using a $U(1)$-index theorem.
We also show how the anomaly is related to the K-theory
classification of Ramond-Ramond fields. In addition we discuss the
M-theory origin of the D6-brane effective action, and illustrate
the general arguments with a concrete example.

\end{abstract}

\end{titlepage}

\pagestyle{plain}
\setcounter{page}{1}
\newcounter{bean}
\baselineskip18pt


\section{Introduction and Summary}

In recent years D6-branes have proved to be a valuable tool for understanding various
aspects of M-theory. For example, one can construct semi-realistic
Standard-like Models in four dimensions from configurations of D6-branes
 -- for a recent example, see \cite{cvetic} and references therein. Such
configurations are also closely related to the subject of M-theory
on conical singularities, where again the dual description in
terms of D6-branes in type IIA has proved useful in understanding
the dynamics -- see, for example, \cite{AW, GS, GST}. From the
M-theory perspective, the D6-brane is a Kaluza-Klein monopole.
Reduction from M-theory to type IIA involves choosing an
``M-theory circle", and the D6-brane is then, roughly speaking, a
codimension four locus $Q$ over which this circle degenerates. In
this way, M-theory is dual to type IIA string theory with
D6-branes wrapped on $Q$. The dynamics of the latter may then
often be understood using standard string theory techniques.

The low energy dynamics of a D6-brane, as for all D-branes, is
governed by a Born-Infeld effection action, together with certain
Wess-Zumino terms which couple the worldvolume fields to the type
IIA Ramond-Ramond fields. For the simple case of a flat D6-brane
linearly embedded in flat Minkowski spacetime, it is fairly
straightforward to derive (much of) this effective action from
Kaluza-Klein reduction of eleven-dimensional supergravity on
Taub-NUT space \cite{im} (for work on the dyonic nature of this
D6-brane, see \cite{spanish1, spanish2}). However, in general it
seems that the precise way in which the D6-brane dynamics arises
from M-theory is not well-understood. This gap was partially
filled in \cite{me} where it was shown how the gravitational
Wess-Zumino terms on a D6-brane arise from the M-theory effective
action -- for M-theory compactified on a spin eight-manifold, the
gravitational couplings on a D6-brane were shown to arise from a
non-standard Kaluza-Klein reduction of a higher-derivative
gravitational correction to the eleven-dimensional supergravity
action, in which bulk couplings reduce to brane couplings.

In this note we address what is perhaps an even more fundamental
question: how does the gauge field on the D6-brane arise from
M-theory? Recall that on every D-brane there propagates a $U(1)$
gauge field $A$, with field strength $F=\diff A$. If we measure
the gauge field in terms of the flux of its field strength,
$[F]\in H^2(Q;\mr)$, where the brackets denote cohomology class,
then we would like to know how $[F]$ is related to the dual
M-theory description. Moreover, the field strength $F$ is Dirac
quantised in the quantum theory, and one should again be able to
see this from M-theory. In fact, there is an interesting subtlety
in this quantisation condition which will be a focal point of this
note. Naively, one expects the periods of $F$ to be integer
multiples of $2\pi$, so that in fact $[F/2\pi] \in
H^2(Q;\mathbb{Z})$ -- this is Dirac quantisation. However, by
studying global worldsheet anomalies for a fundamental string
which ends on a D-brane $Q$, Freed and Witten \cite{freedwitten}
showed that the gauge field strength $F$ on a D-brane should in
general satisfy a modified form of Dirac quantisation
\be
\int_U \frac{F}{2\pi} = \frac{1}{2}\int_U w_2(Q) \quad \mathrm{mod}\
\mz\label{Fshift}\ee
where $U\subset Q$ is any two-cycle on the D-brane $Q$, and
$w_2(Q)$ is the second Stiefel-Whitney class of $Q$. The latter is
non-zero precisely when $Q$ is not a spin manifold. Thus when
spinors exist globally on $Q$, the quantisation condition on $F$
is the naive one.  However, more generally (\ref{Fshift}) says
that the periods of the field strength are shifted to be
half-integer multiples of $2\pi$. In the case of a D6-brane, the
question of how this shift in the quantisation of $F$ is related
to M-theory arose in a specific example in \cite{gomis} where a
D6-brane was wrapped on a supersymmetric cycle $\mcp^2$. This
example was subsequently analysed in considerable detail in
\cite{GS,GST} -- our example in section 4 is a certain
compactification this.

In this paper we explain how $[F]$ is related to the dual M-theory
description in the case where the M-theory background is smooth,
and there is no source for the $G$-flux. We also show precisely
how the Freed-Witten quantisation condition (\ref{Fshift}) arises
in this context. Specifically, we begin with the membrane anomaly
in M-theory and show that it reduces precisely to the Freed-Witten
anomaly described above, in the particular case where the membrane
reduces to a fundamental string ending on a D6-brane. The proof
requires a ``localisation formula'' which we derive using a
$U(1)$-index theorem. We conclude this section with a summary of
our results.

In order to analyse the well-definedness of the string worldsheet
path integral in the presence of D-branes, Freed and Witten
studied one-parameter families of string worldsheets ending on
some D-brane $Q\subset Y$, where $Y$ denotes spacetime. Thus
consider such a loop of string worldsheets $\Sigma \times {\bf
S}^1\subset Y$, with $U\equiv
\partial \Sigma \times {\bf S}^1\subset Q$, and ${\bf S}^1$
parametrises the loop. Since the fundamental string lifts to the
membrane, we obtain a one-parameter family of membranes when we
lift to M-theory. However, the D6-brane is a Kaluza-Klein
monopole, and therefore each membrane $W$ in the family must be
\emph{closed} -- that is, without boundary -- since there are no
M5-branes or boundaries to spacetime by assumption. Thus we obtain
a one-parameter family of closed membranes $V\equiv W \times {\bf
S}^1$.

In general, the M-theory four-form field strength $G$ also
satisfies a shifted quantisation condition, which again is derived
by studying one-parameter families of closed membranes
\cite{wittenflux}. The quantisation condition is
\be
\int_V \frac{G}{2\pi} = \frac{1}{2}\int_V \frac{1}{16\pi^2} \mathrm{tr}\mathcal{R}
\wedge \mathcal{R} \quad \mathrm{mod} \ \mathbb{Z}\label{Gshift}\ee
where $\mathcal{R}$ denotes the curvature two-form for the
M-theory spacetime $X$, and $V$ may be any four-cycle, although
the case of interest for us will be $V=W\times {\bf S}^1$ as
defined above. If $X$ is spin, the quantity inside the integral on
the right hand side of (\ref{Gshift}), which is half the first
Pontryagin form for $X$, is always an integer, but in general is
not divisible by two. Thus the periods of $G$ are shifted to be
half-integer multiples of $2\pi$ in general.

The proof of the Freed-Witten anomaly proceeds in two steps. Firstly, we show that
\be \exp\left(i\int_V G\right) = \exp\left(i\int_U
F\right)\label{GequalsF}\ee
where $U=\partial\Sigma\times {\bf S}^1$ is as defined above. This
follows from a careful analysis of how the gauge field on the
D6-brane arises from the $C$-field in M-theory. In particular, a
crucial physical point to understand here is that, if the M-theory
four-form $G$ is everywhere smooth and closed, then there is no
M5-brane charge in M-theory and thus there is no D4-brane charge
in type IIA. We will see how this physical statement manifests
itself mathematically in a precise way. The second part of the
proof follows from a localisation formula. Specifically, we show
that
\be \int_V \frac{1}{16\pi^2} \mathrm{tr}\mathcal{R} \wedge
\mathcal{R} = \int_U w_2(Q) \quad \mathrm{mod} \
2\label{result}~.\ee
We prove this using a $U(1)$-index theorem for the Dirac operator
on the membrane worldvolume. Putting (\ref{GequalsF}) together
with (\ref{result}) therefore leads to an ``M-theory derivation"
of the Freed-Witten anomaly.

In section 4 we discuss the M-theory origin of the Wess-Zumino
couplings on the D6-brane. The gravitational terms were treated in
\cite{me}. Here we discuss the origin of the gauge field terms.

Since some aspects of this paper are a little technical, we also
include a simple concrete example in section 4. In this case, one
can recover some of the results in this paper by instead analysing
tadpole cancellation. This also ties in naturally with the work of
\cite{me} and the discussion of Wess-Zumino couplings. Finally,
for completeness, we show in section 5 how the Freed-Witten
anomaly also follows from the K-theoretic quantisation condition
for the Ramond-Ramond four-form. This naturally ties in with our
discussion of D4-brane charge. We conclude the paper with some
speculative comments.

As a final comment in this section, notice that the only other
type IIA D-brane for which the Freed-Witten anomaly may be non-trivial is
the D4-brane. In this case the anomaly may be derived from M-theory
by considering the partition function of the chiral two-form
that propagates on the M-theory five-brane \cite{witten}.


\section{Kaluza-Klein Reduction}

As we argued in the introduction, a string ending on a D6-brane must lift to a closed
membrane worldvolume in M-theory. Our aim in the first part of this section
is to describe this more accurately. In particular the discussion here will be
useful in sections 3 and 5. We also give a simple example.

Consider M-theory on an oriented spin manifold $X$. Suppose that
$X$ comes equipped with a circle action\footnote{``circle action"
and ``$U(1)$ action" will be used interchangeably.}, which we will
regard as rotating the ``M-theory circle'' direction. Thus the
orbits of the group action will be the M-theory circle fibres. If
$U(1)$ acts freely -- that is, without any fixed points -- then
M-theory on $X$ is dual to type IIA string theory on the quotient
space $Y=X/U(1)$. This is usual Kaluza-Klein reduction. However,
suppose now that there is a codimension four fixed point set
$Q\subset X$. This is a locus on which the $U(1)$ Killing vector
field vanishes. Then, in this case, the quotient may still be
defined, with the fixed point set being interpreted as a D6-brane
in type IIA. Mathematically we are using the following local
identification
\be
\mr^4/U(1) \cong \mr^3\label{model}\ee
for the normal space to $Q$, where we write
$\mr^4=\mathbb{C}\oplus\mathbb{C}$ and the $U(1)$ acts as
multiplication by $e^{i\theta}$ on \emph{both} factors, where
$0\leq\theta\leq2\pi$ is the $U(1)$ group parameter. The origin in
$\mr^3$, where the D6-brane is located, descends from the fixed
origin on the left hand side of (\ref{model}). This local model of
course describes the reduction of Taub-NUT, which is topologically
$\mr^4$, to $\mr^3$ with a Kaluza-Klein monopole at the centre.
Our type IIA spacetime is then the quotient $Y=X/U(1)$, where we
use the local model (\ref{model}) to define the quotient space
near to the codimension four fixed point set. An equivalent
construction may be described as follows. The circle action on $X$
induces a complex structure on the normal bundle to $Q$ in $X$.
The type IIA manifold, in a neighbourhood of the D6-brane, is
described by taking the projectivisation of this complex normal
bundle. In more detail, for each point $p\in Q$, the normal space
to $Q$ at $p$ is a copy of $\mc^2$, and we simply divide out by
the Hopf map $\mc^2/\mc_*\cong {\mcp}^1$ on each
space\footnote{This is the same map as (\ref{model}), except that
we have, in addition, projected out the radial direction.}. Thus
the projectivisation is a ${\mcp}^1={\bf S}^2$ bundle over $Q$.
The type IIA spacetime, in a neighbourhood of $Q$, is thus
obtained by filling in each two-sphere fibre with a three-disc. In
this construction, the zero section of the disc bundle is the
D6-brane worldvolume.

Consider now a string worldsheet $\Sigma$ ending on a D6-brane
$Q$, so that $\partial\Sigma\subset Q$. As we argued in the
introduction, this configuration lifts to a
 \emph{closed}
membrane worldvolume in M-theory. For example, if
$\Sigma\cong D^2$ is a two-disc,
with the boundary of the two-disc $\partial\Sigma\cong {\bf S}^1\subset Q$,
then the M-theory lift of this worldsheet is a membrane wrapped on
$W\cong {\bf S}^3$. To see this, it is easier to consider the inverse process
of reducing from M-theory. Thus, given a closed membrane worldvolume $W$, we identify
$W/U(1)\cong \Sigma$ as the string worldsheet \cite{townsend}, where
$U(1)$ rotates around the
M-theory circle direction. A codimension two fixed point set then naturally
becomes a
boundary of the worldsheet when we reduce. Locally we are using the
following identification
\be
\mr^2/U(1) \cong [0,\infty)\label{local}\ee
where 0 on the right hand side descends from the fixed origin on the left
hand side. As a simple example, embed $W\cong {\bf S}^3$ as a
sphere of unit norm in $\mathbb{C}^2=\mathbb{C}\oplus\mathbb{C}$, and consider the circle action
which rotates the \emph{first} factor of $\mathbb{C}$. The fixed point set is
then $\{0\}\times \mathbb{C}$ which becomes a copy of ${\bf S}^1$ on
the three-sphere. The quotient is then a two-disc $D^2$ where we have
used the local model (\ref{local}) on the normal space to the fixed points
to yield a quotient space that has a boundary. One can make this
completely explicit by writing the round metric on ${\bf S}^3$ as
\be \diff s^2 = \diff \psi^2 + \cos^2\psi \diff \theta^2 +
\sin^2\psi \diff \phi^2~.\ee
The $U(1)$ Killing vector we reduce on is $\partial/\partial\phi$.
This vanishes at $\psi=0$ -- the locus is a circle, parameterised
by $\theta$. The quotient space is $D^2$ with metric
\be \diff \psi^2 + \cos^2\psi \diff \theta^2~.\ee Here $\theta$ is
the angular coordinate on the disc  and the radial variable is
$0\leq\psi\leq\pi/2$, where $\psi=0$ is the boundary of the disc
and $\psi=\pi/2$ is the origin.

\subsubsection*{D4-brane charge}

Consider type IIA string theory on $Y$, with the NS $B$-field
temporarily set to zero. In the absence of any branes, the Bianchi
identity for the Ramond-Ramond field strengths simply asserts that
they are closed. In particular, $G_2$ and $G_4$ are both closed.
However, in the presence of a D6-brane wrapped on $Q$ one instead
has
\be \diff G_2 = 2\pi\delta_Q\label{magnetic}~.\ee
Equation (\ref{magnetic}) states that the D6-brane is a magnetic
source for the M-theory Kaluza-Klein field strength, or
equivalently Ramond-Ramond two-form, $G_2$. This follows from the
fact that the circle quotient that we took is essentially a
fibre-wise application of the Hopf map ${\bf S}^3\rightarrow {\bf
S}^2$, which precisely describes the Kaluza-Klein monopole.
Indeed, the latter means that
\be \int_{{\bf S}^2} \frac{G_2}{2\pi} = 1\ee
where ${\bf S}^2$ is any two-sphere that links the D6-brane
worldvolume. In (\ref{magnetic}) $\delta_Q$ denotes a three-form
supported on $Q$ which integrates to one over the normal space to
$Q$. If one sets up local coordinates $y_i$, $i=1,2,3$, on the
normal space to $Q$, we have roughly $\delta_Q = \delta({\bf
y})\diff y_1\wedge \diff y_2\wedge \diff y_3$. However, this
formula is only valid locally (or in the case that the normal
bundle to the brane is trivial). More generally we may use the
following standard construction for $\delta_Q$ \cite{bott}, which
was also used in the analysis of M5-brane anomaly cancellation
\cite{FHMM}. In a neighbourhood of $Q$, minus $Q$ itself, there is
always a globally-defined closed two-form $e_2$, known as the
global angular form, which integrates to one over any two-sphere
that links the D6-brane worldvolume $Q$ -- there is an explicit
formula for this form in terms of the connection on the normal
bundle to the brane \cite{FHMM, HMM}. Then $\delta_Q$ may be taken
to be
\be \delta_Q = \diff\rho(r)\cdot e_2\ee
where $\rho$ is any smooth function of the radial direction $r$ which is zero for
$r\geq \epsilon$, for some $\epsilon>0$, and is $-1$ near to $r=0$. It is then easy to
check that $\delta_Q$ is closed, has compact support, and integrates to $1$ over the
normal space to $Q$. However, in this construction notice that the D6-brane charge
has effectively been smeared out to a radius
$\epsilon$ -- this is because $G_2$ is no longer closed inside this radius. For a
truly localised D6-brane, as arises in the Kaluza-Klein reduction described above,
one needs to take a limit in which $\epsilon\rightarrow 0$. In this limit, the function
$\rho(r)$ simply becomes a Dirac delta-function supported at $r=0$.

Let us now consider the Ramond-Ramond four-form, $G_4$. In the presence of a D6-brane, this
is also not closed in general. One instead has as an equation which is of the form
\be \diff G_4 = \delta_Q \wedge F\label{Bianchi}~.\ee
Here $F$ is the gauge field strength on the D6-brane, where a
pull-back to a tubular neighbourhood of the brane is understood in
(\ref{Bianchi}). The right hand side of (\ref{Bianchi}) arises
from the Wess-Zumino couplings on the D6-brane -- indeed,
(\ref{magnetic}) also arises this way. Equation (\ref{Bianchi})
expresses the fact that a non-zero flux of the $U(1)$ gauge field
on the D6-brane induces an effective D4-brane charge
\cite{douglas}. Mathematically we can interpret the right hand
side of (\ref{Bianchi}) as the cohomology class $\mathcal{T}(F)$
where $\mathcal{T}:H^*(Q)\rightarrow H^{*+3}_{\mathrm{cpt}}(Y)$
maps cohomology classes on $Q$ to compactly supported classes in
spacetime $Y$. For those who know about such things, $\mathcal{T}$
is essentially just the Thom isomorphism for the normal bundle
$NQ$ of $Q$ in $Y$, where we identify the normal bundle with a
tubular neighbourhood of $Q$ and extend the isomorphism by zero
outside this neighbourhood. More details may be found in \cite{MM}
where the general case is discussed. Using Stokes' Theorem we may
integrate (\ref{Bianchi}) over the normal space to $Q$ to obtain
\be
\int_{{\bf S}^2}\frac{G_4}{2\pi} = \frac{F}{2\pi}\label{Gint}\ee
where ${\bf S}^2$ is any two-sphere that links the D6-brane $Q$.
If we have smoothed out the charge to a radius $\epsilon$ this
two-sphere should have radius greater than $\epsilon$. Since $G_4$
is closed away from the brane, $G_4$ defines a cohomology class on
the complement of $Q$ in $Y$. Then equation (\ref{Gint}) may be
regarded as a cohomological statement.

Now let us try to lift this to M-theory. Since, for non-zero flux
$F$, we have non-zero D4-brane charge, we also expect an M5-brane
charge in M-theory. However, we are interested in M-theory
configurations in which the four-form is smooth and closed
everywhere. There should be no M5-brane sources present, otherwise
the membrane anomaly calculation, which will be our starting point
for analysing the Freed-Witten anomaly, will not be
valid\footnote{It is an interesting open problem to study
anomalies for membranes ending on M5-branes.}. In fact, one can
see this problem when one tries to lift the flux $G_4$ to
M-theory. Since $B=0$, $G$ is just the pull-back of $G_4$. In the
limit $\epsilon\rightarrow 0$, so that the D4-brane charge is
strictly confined to the D6-brane, the lift $G$ of $G_4$ is
singular at the locus where the M-theory circle vanishes. This is
hardly surprising -- we see from (\ref{Gint}) that $G_4$ is also
singular on $Q$. We could consider smoothing out the charge to a
radius $\epsilon>0$. This solves the singularity problem, but
since we do not want an M5-brane charge in M-theory, this is not
the right way to proceed.

\subsubsection*{Reduction of $G$ and incorporation of the NS field}

There is an obvious way to cancel the D4-brane charge, but still
have non-trivial gauge field strength $F$. In general, $F$ is
replaced by the gauge-invariant quantity $(F-B)$ in the
Wess-Zumino couplings on the D6-brane. Since $B$ is a potential,
it does not satisfy any quantisation condition, and we may simply
choose $B$ to cancel the flux $F$. Indeed, this is essentially
what happens in the case of configurations which are dual to
smooth M-theory solutions, as we now describe.

Let us begin with some M-theory configuration with flux $G$,
equipped with a circle action with codimension four fixed point
set $Q$. The $U(1)$ action allows us to write
\be G = \tilde{G_4} + H \wedge e_1\label{Gansatz}\ee
where $e_1 = (\diff\psi - C_1)/2\pi$ denotes the global angular
form on the M-theory circle bundle -- that is, $\psi$ is an
angular coordinate on the M-theory circle and $C_1$ is the
(pull-back of the) connection. The $U(1)$ Killing vector is
therefore $\partial/\partial\psi$. Since $G$ is assumed to have no
sources and $\mathcal{L}_{\partial/\partial\psi} G = 0$, where
$\mathcal{L}$ denotes the Lie derivative, it follows that $H$ is
closed. Of course, this is just the statement that there are no
NS5-brane sources present. From (\ref{Gansatz}) $H$ must also be
zero on the locus $Q$ where $\partial/\partial\psi$ vanishes in
order that $G$ be smooth there. This follows since $e_1$ is
singular at $Q$ -- the dual Killing vector field vanishes there.
The fact that $G$ is closed then implies the Bianchi identity
$\diff\tilde{G}_4 = -H\wedge \frac{G_2}{2\pi}$. Notice that,
although $G_2$ is ill-defined on $Q$, $H$ vanishes on $Q$ and
therefore the Bianchi identity is in fact everywhere smooth.

On the other hand, if we incorporate the NS $B$-field in string
theory, equation (\ref{Bianchi}) is modified to read
\be \diff\tilde{G}_4 = -H\wedge \frac{G_2}{2\pi} + \delta_Q \wedge
(F-B)\label{modbianchi}~.\ee
Here $\tilde{G}_4$ is the gauge-invariant four-form of type IIA,
and is identified with what we called $\tilde{G}_4$ in M-theory.
The first term in (\ref{modbianchi}) arises from a bulk
Chern-Simons coupling in type IIA supergravity. On the other hand,
the last term in (\ref{modbianchi}) arises by replacing $F$ by the
gauge-invariant quantity $F-B$ in the Wess-Zumino couplings on the
D6-brane. Clearly, the two Bianchi identities for $\tilde{G}_4$
agree only if $F=B$ on the D6-brane. This is inevitable since, by
assumption, $\tilde{G}_4$ in M-theory is everywhere smooth. In
fact, all that we will need is that the cohomology classes of $F$
and $B$ on $Q$ agree, $[F]=[B]$.

At this point, the reader may notice the following problem. The
potential $B$ may be computed from the $G$-flux in M-theory by
$\diff B=H$. In general this equation is only valid locally since
the cohomology class of $H$ may be non-trivial. Indeed, globally,
$(B,H)$ is really a Cheeger-Simons differential character, as we
discuss at the end of this section. However, $\diff B=H$ certainly
holds in a tubular neighbourhood $T$ of $Q$ in $Y$ since $H$
actually vanishes on $Q$, and $H^3(T)\cong H^3(Q)$ since $Q$ is a
deformation retract of $T$. But then $B$ is only uniquely defined
modulo $B\rightarrow B+a$, where $a$ is a closed two-form on $Y$,
which defines a class $[a]\in H^2(Y;\mathbb{R})$. Thus $[F]_Q$,
where the subscript emphasises that the cohomology is that of $Q$,
is determined only modulo classes on $Q$ that are the restrictions
of cohomology classes on the whole of spacetime, $Y$. If
$[a/2\pi]\in H^2(Y;\mathbb{Z})$ then $B\rightarrow B+a$ is a large
gauge transformation of the $B$-field, and the corresponding
ambiguity in $F$ merely reflects the fact that it is $F-B$ which
is the gauge-invariant quantity on $Q$. However, in general it
seems that one must specify more precisely\footnote{{\it i.e.} not
just the cohomology class of its curvature.} the M-theory
$C$-field in order to obtain $[F]$. As we will see, this ambiguity
in $[F]$ will be irrelevant for deriving the Freed-Witten anomaly,
and also the higher order Wess-Zumino terms on the D6-brane. This
is just as well, since neither of these depends on the choice of
$C$-field which satisfies $[\diff C]=[G]$.

\

Consider now some closed $U(1)$-invariant four-dimensional
submanifold $V\subset X$, with $V$ having a codimension two fixed
point set $U$ on the locus $Q$. The situation of interest is when
$V$ describes a one-parameter family of membrane worldvolumes in
spacetime, which descend to a family of open strings in type IIA
with boundary $U\subset Q$. Consider the integral $\int_V G$. We
compute
\be
\int_V G = \int_{V/U(1)} H\ee
where we have integrated over the M-theory circle. Notice that
$\tilde{G_4}$ has no support over the M-theory circle, and
therefore does not contribute to the integral. In the case that
$H=\diff B$ holds globally on $Y$ we can use Stokes' theorem to
write
\be \int_V G = \int_{V/U(1)} H = \int_U B = \int_U
F\label{back}~.\ee
This implies the formula (\ref{GequalsF}) that we were looking
for. However, recall that $[F]_Q$ is only determined modulo
$i^*[a]$, where $i:Q\rightarrow Y$ denotes the embedding map, and
$[a]\in H^2(Y;\mathbb{R})$. However, for any such $a$, $\int_U a
=0$, since the homology class of $U$ in $Y$ is trivial. This
follows because, by definition, $U$ bounds $V/U(1)$ in $Y$.

To complete the argument we must consider the case when $[H]\neq0
\in H^3(Y)$. In fact, $[H/2\pi]\in H^3(Y;\mathbb{Z})$ as the NS
field strength $H$ is quantised. In this case, $B$ cannot be a
globally defined two-form -- in fact it is more like a $U(1)$
gauge field that has non-trivial first Chern class. Technically
this means that $(B,H)$ is a ``Cheeger-Simons differential
character". Concretely, this means that, due to the fact that $B$
is not a globally defined object, the integral of $B$ over a
topologically trivial two-cycle $U\subset Y$ is only defined
modulo $2\pi $. Thus
\be \int_U B = \int_{V/U(1)} H \quad \mathrm{mod} \ 2\pi\ee
where recall that $U$ is the boundary of $V/U(1)$ in the case at
hand. The idea here is that, if $Z_1, Z_2$ are any two 3-cycles
with boundary $\partial Z_i = U$ ($i=1,2$) then the integrals
\be \int_{Z_i} H\ee
differ by $\int_Z H = 0$ mod $2\pi$, where $Z$ is the
\emph{closed} three-manifold obtained by gluing $Z_1$ to $Z_2$
(with opposite orientation) along their common boundary $U$, so $Z
= Z_1\cup_U (-Z_2)$. Thus the definition
\be \int_U B = \int_{Z_i} H \quad \mathrm{mod} \ 2\pi\ee
is well-defined -- {\it i.e.} independent of the choice of $i=1,2$
-- and of course is certainly true when $H$ is exact. Thus more
generally (\ref{back}) holds modulo $2\pi$. Thus, exponentiating
everything (multiplied by $i$), we have proved (\ref{GequalsF}).

In section 4 we will examine a concrete example and compute explicitly some of the
quantities appearing in this section. In particular, $H$ is topologically trivial
on $Y$ in this case, and there is a completely
independent check on (\ref{back}) from tadpole cancellation.


\section{Fermion Anomalies and a Localisation Formula}
\label{pffaf}

Our aim in this section is to give a proof of the ``localisation formula''
(\ref{result}).

\subsubsection*{A review of the membrane anomaly}

Consider the worldvolume theory for a membrane propagating on
an oriented spin manifold $X$. Let $W$ be a closed three-dimensional
submanifold\footnote{In fact, more generally we may allow $W$ to be an
immersion.} of $X$, and consider wrapping the membrane on $W$. We focus
on the following two terms in the membrane effective theory
\be \Pf(\D_W)\cdot\exp\left(i\int_W
C\right)\label{worldvolume}~.\ee
Here $\Pf(\D_W)$ denotes the Pfaffian of the Dirac operator on the
membrane worldvolume, and the second term in (\ref{worldvolume})
is the ``holonomy'' of the M-theory $C$-field over $W$. In order
to describe the Pfaffian, let $N_W$ denote the normal bundle of
$W$ in $X$. This has rank eight, and, since $W$ and $X$ are both
spin, there exists a spin covering, which we denote
$S(N_W)=S_+(N_W)\oplus S_-(N_W)$, and we have decomposed into
positive and negative chirality. The membrane worldvolume
fermions, after fixing kappa symmetry, are then spinors on $W$
with values in $S_+(N_W)$ -- also a real bundle of rank eight.
Now, the spinors on $W$ are pseudoreal. The Dirac operator is
Hermitian, and therefore its eigenvalues are real. For fermions
valued in any real bundle, for example $S_+(N_W)$, the eigenvalues
come in pairs -- this is due to the existence of an anti-unitary
symmetry in three dimensions. The fermion path integral is then
given by the Pfaffian of the Dirac operator valued in $S_+(NW)$,
$\Pf(\D_W)$, and is roughly the square root of the determinant.
Since the eigenvalues come in pairs, the determinant is formally
positive, and regularisation preserves this property. The Pfaffian
is then formally given by
\be
\Pf(\D_W) = \prod_n \lambda_n\label{pfaffian}\ee
where the product runs over \emph{pairs} of eigenvalues -- that
is, we include the contribution from one of the eigenvalues in
each pair. The (regularised) Pfaffian is then real, but there is
no natural definition of its sign. This can potentially lead to an
anomaly. Indeed, if one deforms the membrane around a
one-parameter loop, the spectral flow in (\ref{pfaffian}) might
mean that the Pfaffian changes sign as one traverses the loop.
This spectral flow is given by the topological index theorem for
families of Dirac operators. In order to describe this, notice
that
\be
N_W = \mathcal{O}\oplus N^{\prime}\ee
where $\mathcal{O}$ is a trivial real line tangent to the ${\bf
S}^1$ which parameterises the family of worldvolumes. Thus
$N^{\prime}$ is a $Spin(7)$ bundle. The number of eigenvalue pairs
of the Dirac operator $\D_W$ that change sign in going around the
circle (the spectral flow) is then given by \cite{wittenflux}
\be \frac{1}{2}\Index\  \D_{W\times {\bf S}^1}\ee
where $\D_{W\times {\bf S}^1}$ denotes the chiral Dirac operator
on $W\times {\bf S}^1$, coupled to $S(N_{W\times{\bf S}^1})$. This
operator arises by essentially gluing together the Dirac operators
on $W\times \{q \}$, for each $q\in{\bf S}^1$, to make a Dirac
operator on $W\times {\bf S}^1$. Note that, in four dimensions,
the Dirac operator coupled to any real vector bundle has an index
which is divisible by two and thus the above expression is indeed
an integer. Using the index theorem one finds
\be \frac{1}{2}\Index\  \D_{W\times {\bf S}^1} = \int_{W\times
{\bf S}^1} \frac{p_1(X)}{2} \quad \mathrm{mod} \
2~.\label{index}\ee
In order to obtain this result one needs to note that all the
characteristic classes of the tangent bundle of $W\times {\bf
S}^1$ vanish.

To conclude, in order that the membrane quantum effective action be well-defined, the change in
the sign of the Pfaffian in (\ref{worldvolume})
as one traverses the loop must be cancelled by the change in the holonomy
factor $\int_W C$. This leads to the non-standard Dirac quantisation condition (\ref{Gshift}) for
$G$.

\subsubsection*{A $U(1)$-index calculation}

The idea in this section is simply to apply the $U(1)$-index
theorem, where $U(1)$ acts by rotating the M-theory circle fibres,
to the Dirac operator $\D_{W\times {\bf S}^1}$, instead of the
usual index theorem. This will lead directly to the result
(\ref{result}). We will therefore need to describe what the
$G$-index theorem is.

It might be useful to first recall some of the details of the
usual index theorem. We begin with a sequence $\{\E_k\}$ of smooth
vector bundles over $M$, labelled by $k$, only finitely many of
which are non-zero. The space of smooth sections of $\E_k$ is
denoted $\Gamma(\E_k)$. We also assume we have differential
operators $\D_k: \Gamma(\E_k) \rightarrow \Gamma(\E_{k+1})$ which
form a \emph{complex}: $\D_{k+1}\circ \D_k = 0$. If
$\D_k^*:\Gamma(\E_{k+1})\rightarrow \Gamma(\E_k)$ is the dual
operator, and $\Delta_k \equiv \D_k^*\D_k + \D_{k-1}\D^*_{k-1}$ is
the Laplacian, then the complex is said to be \emph{elliptic} if
the Laplacian is an elliptic operator on $\Gamma(\E_k)$. We define
the cohomology of the elliptic complex, in the usual way, to be
\be H^k(\E,\D) = \frac{\ker(\D_k)}{\mathrm{im} (\D_{k-1})} = \ker
\Delta_k\ee
where the second equality follows from Hodge-de Rham theory, just
as for the de Rham cohomology of manifolds -- indeed the latter is
just a special case of the above where $\E_k=\Lambda^k$ is the
bundle of $k$-forms and the operators $\D_k$ are just the exterior
derivative $\mathrm{d}$, restricted to $\Lambda^k$. The
\emph{index} of the elliptic complex $(\E,\D)$ is then defined to
be
\be \Index (\E,\D) = \sum_k (-1)^k \dim H^k(\E,\D)~.\ee
For example, for the exterior algebra this is just the Euler
number of the base manifold $M$. The celebrated index theorem of
Atiyah and Singer \cite{AS} relates the index, which is an
analytic object, to certain characteristic classes integrated over
$M$, which is a purely topological object.

Quite generally, we may also assume that we are given an action of
the group $G$ on our complex. Thus for each group element $g\in G$
we have smooth bundle maps $g_k:\E_k\rightarrow \E_k$ which
commute with the $\D_k$ operators, $g_{k+1}\circ \D_k =
\D_{k}\circ g_k$, and so lead to an induced action on the
cohomology groups, $\hat{g}_k: H^k(\E,\D)\rightarrow H^k(\E,\D)$.
We may then define the \emph{Lefschetz number} to be
\be L(g,\E,\D) = \sum_k (-1)^k \mathrm{tr}
(\hat{g}_k)~.\label{lef}\ee
Clearly, if the action of $g$ is trivial, this is just the usual
index since the trace just computes the dimension of the
cohomology groups. More generally we get a character of $g$. The
$G$-index theorem is then a generalisation of the usual index
theorem in which there is, in addition to the elliptic complex, a
specified action of the group\footnote{There are some additional
technical assumptions that we are suppressing for the time being.
For example, $G$ must be ``topologically cyclic", which simply
means that there is a group element $g\in G$ whose powers are
dense in $G$.} $G:M\rightarrow M$, which ``lifts" to an action on
the complex $(\E,\D)$ as described above \cite{AS}. The theorem
then computes the Lefschetz number in terms of certain cohomology
classes evaluated on the \emph{fixed point set} $M^g$ of $g$ in
$M$. Again, if the action of $g$ is trivial, the theorem reduces
to the usual index theorem.

To return to our problem, we have $G\cong U(1)$ which acts on the
loop of membrane worldvolumes $W\times {\bf S}^1$. This will be
the base $M$. We then apply the $U(1)$-index theorem to the Dirac
operator $\D_{W\times {\bf S}^1}$. This will give a formula for
the Lefschetz number (\ref{lef}) in terms of certain
characteristic classes evaluated on the fixed point set
$U=\partial\Sigma\times {\bf S}^1$, which recall physically is a
loop of string worldsheet boundaries on the D6-brane. Thus in the
$U(1)$-index theorem we have $U = M^g$, for non-trivial $g\in
U(1)$. In order to utilise the theorem, we will need to work out
how $U(1)$ acts on the relevant bundle, which is essentially the
spin bundle associated to the normal bundle $N_{W\times {\bf
S}^1}$ of $W\times {\bf S}^1$ in $X$. This action is of course
induced from the action of $U(1)$ on the embedding space, $X$ --
the M-theory manifold in which the loop of membranes is sitting.
As we will see explicitly below, the Lefschetz number is in fact
\emph{independent} of the particular group element $g \in U(1)$
chosen, and is thus, setting $g$ equal to the identity, equal to
the usual index. The upshot is thus a formula for the usual index
in terms of certain cohomology classes evaluated on the fixed
point set of the circle action. Of course, this is precisely the
form of (\ref{result}). Before we begin with the details, we
suggest here that the interested reader might consult section 7.6
of \cite{EGH} for further background on $G$-index theorems, as
well as some simpler examples.

We denote the bundle $S(N_{W\times{\bf S}^1})$ as $E$, and the
Dirac operator on $W\times {\bf S}^1$ coupled to $E$ as $\D$, and
of course $W\times {\bf S}^1$ is $M$. Thus we have a two-term
complex $\E = S^+\otimes E, S^-\otimes E$, where $S^{\pm}$ are the
chiral spinor bundles of the four-manifold $M=W\times {\bf S}^1$.
We are now in line with the notation above. The membrane anomaly
is computed from $\mathrm{Index}(\E,\D)$, and we would now like to
compute the Lefschetz number by applying the $G$-index theorem.

It is a relatively straightforward exercise to write down the form
of the $G$-index theorem for a Dirac operator coupled to a vector
bundle $E\rightarrow M$ with $G$-action, starting from the general
theorem in \cite{AS}. The result is
\be L(g, \E, \D) = (-1)^l \int_{M^g} \prod_{j} \left(-2i \sin
\frac{\theta_j}{2}\right)^{-s_j} \hat{A}(M^g)\ \prod_{j}
\mathcal{M}^{\theta_j}(N^g(\theta_j)) \
\mathrm{ch}(E\mid_{M^g})(g)~.\label{Gindex}\ee
This rather formidable expression is fairly straightforward to
explain. $M^g$ denotes the fixed point set of $g$, which has
dimension $2l=\dim M^g$, and $N^g$ is its normal bundle in $M$.
Now, since $g$ fixes $M^g$ we get an induced action on its normal
bundle $N^g$ in $M$, which then induces a splitting of $N^g$ into
a sum complex vector bundles and real vector bundles. This follows
from simple representation theory of cyclic groups -- recall that
$G$ is required to be topologically cyclic. In fact we have
omitted the contribution from the real vector bundle in
(\ref{Gindex}) since it is not present for a generic element of
$U(1)$ -- for $G\cong U(1)$, we already remarked earlier in
section 2 that we obtain an induced complex structure on the
normal bundle. Thus we have a bundle decomposition
\be N^g = \bigoplus_{j} N^g(\theta_j)~.\ee
where the action of $g$ on the complex vector bundle
$N^g(\theta_j)$ is by definition multiplication by
$e^{i\theta_j}$. Moreover, the complex dimension of this bundle is
$s_j$, so that $\sum_j s_j = s = \dim_{\mathbb{C}} N^g$.

Finally, $\hat{A}$ is the usual Dirac genus,
\be
\hat{A} = 1 - \frac{p_1}{24} + \frac{7p_1^2-4p_2}{5760}+\ldots\ee
$\mathcal{M}^{\theta}$ is the (stable) characteristic class given
by the formula\footnote{This formula was also used in \cite{me}.}
\be \mathcal{M}^{\theta} = 1 + \frac{i}{2}\cot \frac{\theta}{2}
c_1 + \ldots\ee
where the dots denote higher order terms which we will not need,
and $\mathrm{ch}(E\mid_{M^g})(g)$ denotes the equivariant Chern
character of the bundle $E$ restricted to the fixed point set
$M^g$. This Chern character replaces the usual Chern character in
the index theorem, and is in fact the most important term for us.
We will therefore need to describe this object.

Recall that, in the usual index theorem, one encounters the Chern
character $\mathrm{ch}(E) \in H^*(M;\mathbb{Q})$. For example, for
a spin bundle $E$ associated to a vector bundle $N$ -- the case of
interest here -- one has the formula
\be \mathrm{ch}(E) = \prod_k
\left(e^{y_k/2}+e^{-y_k/2}\right)~.\label{spin}\ee
where the $y_k$ are the basic characters of $N$. If one
``imagines" that $N$ is in fact a direct sum of complex line
bundles, then the $y_k$ are just the first Chern classes of these
line bundles. This is known as the splitting principle.

In the equivariant case we encounter the object
$\mathrm{ch}(E\mid_{M^g})(g)$. The $G$-index theorem naturally
involves \emph{equivariant} K-theory, which is where this object
comes from. Here we simply sketch the ideas and state the formula
-- for further details the inquisitive reader should consult the
literature.

Consider the situation where the group $G$ acts trivially on a
manifold $Z$, which for us will be the fixed point set $M^g$ in
the $G$-index theorem. Then a basic result is that the equivariant
K-theory of $Z$ is simply the tensor product
\be K_G(Z)\cong K(Z)\otimes R(G)\ee
where $R(G)$ denotes the character ring of $G$. Suppose then that
we have an element $u = x\otimes \chi\in K_G(Z)$ where $x\in
K(Z)$, and $\chi$ is a character of $G$. Then the equivariant
Chern character \cite{AS} is a map from $K_G(Z)\rightarrow
H^*(Z;\mathbb{C})$ defined as follows:
\be \mathrm{ch} (u)(g) = \chi(g)\cdot \mathrm{ch}(x)\in
H^*(Z;\mathbb{C})\ee
where $\mathrm{ch}(x)$ is the usual Chern character. In the
$G$-index theorem we encounter the complex vector bundles
$N^g(\theta_j)$ where, by definition, the action of $g\in G$ on
$N^g(\theta_j)$ is multiplication by $e^{i\theta_j}$. If $y_k$
denote the basic characters of $N^g(\theta_j)$,
$k=1,\ldots,s_j=\dim_{\mathbb{C}} N^g(\theta_j)$, we have, for
fixed $j$, the
 contribution
\be \mathrm{ch}(u)(g) = e^{i\theta_j} \cdot
\mathrm{ch}(N^g(\theta_j)) = \sum_k e^{i\theta_j}\cdot e^{y_k} =
\sum_k e^{y_k+i\theta_j}~.\ee
Thus, in the equivariant Chern character, one simply replaces the
basic characters $y_k$ of the $j$'th complex vector bundle by
$y_k+i\theta_j$ in the formula for the usual Chern character. We
will use this fact below.

We now specialise to the case $G\cong U(1)$. The fixed point set
is $M^g = U = \partial\Sigma \times {\bf S}^1$, and so $l=1$, and
the normal bundle of $U$ in $M=V=W\times{\bf S}^1$ is just a
complex line, so that $s=1$. Indeed, notice that the normal bundle
to $Q$ in $X$, which we denote by $\mathcal{V}$, has real rank
four. $\mathcal{V}$ therefore splits into the sum of two complex
line bundles under the $U(1)$ action when restricted to $U$,
$\mathcal{V}\mid_U = \mathcal{L}_1\oplus \mathcal{L}_2$, with
$U(1)$ acting as multiplication by $e^{i\theta}$ on each factor --
here $0\leq\theta\leq2\pi$ is now literally the $U(1)$ group
parameter $g=e^{i\theta}$. We may take $\mathcal{L}_1$ to be
normal to $\partial\Sigma\times {\bf S}^1$ in $W\times {\bf S}^1$
-- that is, $\mathcal{L}_1$ is just the normal bundle $N^g$ of
$M^g$ in $M$. Thus $N^g=N^g(\theta)$, in the above notation.

Evaluating (\ref{Gindex}) we obtain
\be L(\theta, \E, \D) = -\frac{i}{2}\cosec \frac{\theta}{2}
\int_{U} \left[1+
\frac{i}{2}\cot\frac{\theta}{2}c_1(\mathcal{L}_1)+\ldots\right]\
\mathrm{ch}(E\mid_U)(g)\label{halfwayhouse}~.\ee
It remains to compute the equivariant Chern character. Recall that
$E$ is the spin bundle for $N_M$, where $M=W\times{\bf S}^1$.
$N_M$ is a rank seven vector bundle (since $7=11-4$), and over the
fixed point set $U$ we obtain a splitting
\be
N_M\mid_{U} = \mathcal{F}\oplus \mathcal{L}_2\ee
where $\mathcal{L}_2$ is the line bundle which appears in the
decomposition $\mathcal{V}\mid_U = \mathcal{L}_1\oplus
\mathcal{L}_2$. The group $U(1)$ acts trivially on the real rank
five bundle $\mathcal{F}$ -- which corresponds to the directions
in the D6-brane transverse to the loop of string boundary -- but
rotates the line bundle $\mathcal{L}_2$ by the action
$e^{i\theta}$. Let us denote the basic characters for
$\mathcal{F}$ as $y_k$, $k=1,2$, and let the first Chern class of
$\mathcal{L}_2$ be denoted $y=c_1(\mathcal{L}_2)$. Recalling that
$E$ is the spin bundle associated to $N_M$, we may therefore
compute the equivariant Chern character using (\ref{spin}):
\be \mathrm{ch}(E\mid_U)(\theta) =
\prod_k\left(e^{y_k/2}+e^{-y_k/2}\right)
\cdot\left(e^{(y+i\theta)/2}+e^{-(y+i\theta)/2}\right)~.\ee
This follows from our above discussion, where we argued that the
Chern class $y$ of $\mathcal{L}_2$ gets replaced by $y+i\theta$ in
the equivariant formula; $\mathcal{F}$ has a trivial $U(1)$
action, and so the $\theta_k=0$ for this bundle.

On substituting this expression into (\ref{halfwayhouse}) and using
standard trigonometric formulae, one arrives at the result
\be L(\theta, \E, \D) = \int_{U} 2c_1(\mathcal{L}_2) +
2\cot^2\frac{\theta}{2}c_1(\mathcal{L}_1)~. \ee
Now, in fact
\be \int_{U} c_1(\mathcal{L}_1) = 0\label{sumup}~.\ee
This is a simple consequence of the fact that $\mathcal{L}_1$ is
the normal bundle of $M^g=U$ in $M=V$, and the circle action which
rotates $\mathcal{L}_1$ extends over $M=V$ without any other fixed
points. Specifically, we have
\be \int_U c_1(\mathcal{L}_1) = \int_{V/U(1)}
\mathrm{d}\left[c_1(\mathcal{L}_1)\right] = 0\ee
where $V/U(1)$ is the quotient three-manifold, which has boundary
$U$, and we have simply used Stokes' theorem. In general, $M^g=U$
need not be connected and then, in principle, $c_1(\mathcal{L}_1)$
could be non-zero on some components, as long as (\ref{sumup})
holds. Thus we see that the Lefschetz number is actually
\emph{independent} of $g=e^{i\theta}$. But setting $g=1$ in
(\ref{lef}) we of course obtain the usual index. Thus we have
shown that
\be \frac{1}{2}\Index\ \D = \int_{U}
c_1(\mathcal{L}_2)\label{indexchar}~.\ee
Now, we have $y=c_1(\mathcal{V})$, restricted to $U$, where we
have used (\ref{sumup}). But then
$c_1(\mathcal{V})=w_2(\mathcal{V})$ modulo two, and since $X$ is
oriented and spin it follows that $c_1(\mathcal{V})=w_2(Q)$ modulo
two also. Thus
\be \frac{1}{2}\Index\ \D = \int_U w_2(Q) \quad \mathrm{mod} \
2~.\ee
Combining this with the usual index theorem (\ref{index}) gives
(\ref{result}).


\section{Wess-Zumino Couplings and an Example}

In this section we consider a concrete example where the anomalies
in question are non-trivial, and also discuss the Wess-Zumino
terms on the D6-brane. There is then an independent check of some
of our results which arises by considering tadpole cancellation.
This also ties in naturally with reference \cite{me}.

Consider M-theory on $\mr^{1,2}\times {\mhp}^2$, where ${\mhp}^2$
denotes quaternionic projective two-space. The isometry group of
this space is $Sp(3)$ and there is an embedding $U(3)\subset
Sp(3)$, which amounts to the embedding $\mathbb{C}\subset
\mathbb{H}$. Then the action of the diagonal $U(1)\subset U(3)$
on\footnote{In this section the $\mr^{1,2}$ factor does not play
any role in the discussion, and so we project it out of our
formulae. Thus $X$ will denote the non-trivial part of the
M-theory spacetime. Similar remarks will apply elsewhere.} $X =
{\mhp}^2$ has a fixed point set ${\mcp}^2$ \cite{AW, GST}. In
fact, the generic orbit under the $U(3)$ action is a copy of the
Aloff-Wallach space $N_{1,-1}=SU(3)/U(1)$. This has codimension
one in $X$. There is then a theorem that we may apply which which
states that there are then precisely two ``special orbits" of
higher codimension. In the case at hand, one of these is a copy of
${\bf S}^5$, which is Hopf-fibred over ${\mcp}^2$ by the circle
action, and the other is a copy of ${\mcp}^2$, which is left fixed
by $U(1)$. The latter is thus a codimension four fixed point set
and so will become our D6-brane worldvolume $Q$. A full discussion
of this orbit structure may be found in \cite{AW}.

Since ${\mcp}^2$ is not spin, the gauge field strength on a
D6-brane wrapped on $\mr^{1,2}\times {\mcp}^2$ has periods which
are half-integer multiples of $2\pi$. In fact, there is only one
non-trivial two-cycle $U={\mcp}^1\subset {\mcp}^2$, and so we may
generally write
\be
\int_{{\mcp}^1} \frac{F}{2\pi} = \frac{1}{2}+n\ee
for some $n\in\mathbb{Z}$. This integer completely characterises the flux in this
case.

It is also easy to analyse the membrane anomaly on $X={\mhp}^2$.
The integral cohomology $H^4({\mhp}^2;\mathbb{Z})$ is generated by
a four-form $\lambda$. In fact it is quite straightforward to show
that $\lambda$ is precisely half the first Pontryagin class of
${\mhp}^2$. It follows of course that $\lambda$ is not divisible
by two, and so the membrane anomaly is non-trivial -- indeed, from
the last section we know that this must be the case. As for the
complex projective space, ${\mhp}^2$ has only one non-trivial
cycle -- dual to $\lambda$ -- which is a linearly embedded
${\mhp}^1\cong {\bf S}^4$. In general we may therefore write
\be
\int_{{\mhp}^1} \frac{G}{2\pi} = \frac{1}{2}+m\ee
for some $m\in\mathbb{Z}$, which again completely characterises
the four-form flux.

Now, the cycle ${\mhp}^1$ is acted on by the circle action, with
fixed point set being the non-trivial ${\mcp}^1\subset {\mcp}^2$.
Indeed, notice that the circle action on ${\mhp}^1$ \emph{must}
have fixed points somewhere since the Euler number of ${\mhp}^1$
is $2$. The Lefschetz fixed point formula then asserts that the
Euler number is the sum of the Euler numbers of the fixed point
sets -- this is in fact a simple application of the $G$-index
theorem where one uses the de Rham complex. Thus there are only
two obvious possibilities -- either one fixes two points (the
north and south poles of the four-sphere), or else one fixes a
copy of ${\bf S}^2$. In fact it is the latter that is the case
here. For further details the reader is referred to \cite{GST}.
Moreover, the quotient space $Y=X/U(1)$ is extremely simple -- it
is the seven-sphere ${\bf S}^7$. This non-obvious fact is proved
in \cite{AW}. In particular, notice that $H$ is necessarily
cohomologically trivial since $H^3({\bf S}^7;\mathbb{Z})=0$ and
thus our formula (\ref{back}) holds and gives $n=m$. We next
proceed to show how one may use tadpole cancellation to check
this. However, before doing this, it might be instructive to
explicitly compute the terms in the $G$-index theorem for this
space.

The index of the Dirac operator on ${\bf S}^4\subset X$ coupled to
the normal spin bundle gives
\be \frac{1}{2}\Index \ \D = \int_{{\bf S}^4} \lambda = 1~.\ee
where the last step follows since $\lambda$ is dual to the
four-cycle ${\bf S}^4$, as discussed above. According to our
$U(1)$-index calculation (\ref{indexchar}) this should equal
\be \int_{{\bf S}^2} c_1(\mathcal{V})\label{inty}\ee
where recall that $\mathcal{V}$ is the normal bundle of ${\mcp}^2$
in $X$ -- {\it i.e.} the normal bundle of the fixed point set.
Thus we need to know the normal bundle of ${\mcp}^2$ in
${\mhp}^2$. In fact, as shown\footnote{This example is closely
related to certain $Spin(7)$ manifolds.} in \cite{GST}, this is
the ``universal quotient bundle" of ${\mcp}^2$. This may be
defined as follows. One begins with the trivial bundle over
${\mcp}^2$ of complex rank 3, which is thus simply the product
${\mcp}^2\times \mathbb{C}^3$ . There is then a natural complex
line bundle $\mathcal{S}$ over ${\mcp}^2$ which may be defined as
the subbundle consisting of pairs $(p,l)\in {\mcp}^2\times
\mathbb{C}^3$ where $l$ is the complex line in $\mathbb{C}^3$
corresponding to the point $p\in {\mcp}^2$. The universal quotient
bundle is then simply the orthogonal complement of $\mathcal{S}$
in ${\mcp}^2\times \mathbb{C}^3$. From this definition one easily
sees that the first Chern class of this rank two complex vector
bundle is indeed equal to the generator of
$H^2({\mcp}^2;\mathbb{Z})\cong \mathbb{Z}$, which is dual to the
two-cycle ${\bf S}^2$, so that the integral (\ref{inty}) is 1.
Thus we have verified the result of our general $U(1)$-index
calculation in this explicit example.

\subsubsection*{Wess-Zumino terms and tadpole cancellation}

As reviewed in \cite{me}, tadpole cancellation for the $C$-field yields
\be N_{M2} + \frac{1}{192}\int_X \left(p_1^2-4p_2\right) +
\frac{1}{2}\int_X \left( \frac{G}{2\pi}\right)^2 =
0\label{Ctad}~.\ee
Here $N_{M2}$ is the number of space-filling M2-branes in
$\mr^{1,2}$, located at some points in $X$, and $p_i$ denote
Pontryagin classes. The latter arise from the gravitational
correction to the eleven-dimensional supergravity action mentioned
in the introduction. The last term in (\ref{Ctad}) comes from the
usual Chern-Simons term in eleven-dimensional supergravity.

There is a similar tadpole condition for $C_3$ in type IIA which
arises due to the Wess-Zumino terms on a D6-brane \cite{me}
wrapped on $\mr^{1,2}\times Q$. In general there is also a bulk
contribution. The tadpole condition then reads
\be N_{D2} + \int_Q \sqrt{\hat{A}(TQ)/\hat{A}(NQ)} +
\frac{1}{2}\int_Q \left( \frac{F-B}{2\pi}\right)^2 + \int_Y
\frac{\tilde{G}_4}{2\pi}\wedge \frac{H}{2\pi} =
0\label{C3tad}~.\ee
Here $N_{D2}$ is the number of space-filling D2-branes, which is
clearly identified with $N_{M2}$ in M-theory.

The main result of \cite{me} was that, for general $X$ and
D6-brane configuration $Q$, the gravitational terms in
(\ref{Ctad}) and (\ref{C3tad}) are equal. Using the results of
section 2, one can verify that the remaining Wess-Zumino and bulk
terms in (\ref{C3tad}) descend from the $G$-flux terms in
(\ref{Ctad}). Indeed, the Wess-Zumino term involving $F-B$ of
course vanishes. Since the bulk term in (\ref{C3tad}) is simply
the dimensional reduction of the flux term in (\ref{Ctad}), the
equivalence of the tadpole conditions is then clear.

However, one may now integrate the bulk term in (\ref{C3tad}) by
parts\footnote{For simplicity we assume that $H$ is topologically
trivial. In particular this is the case in our example.} and,
using the Bianchi identity for $\tilde{G}_4$ and fact that $F=B$
on $Q$, one finds that the bulk term may be written as
\be \frac{1}{2}\int_Q
\left(\frac{F}{2\pi}\right)^2\label{crap}~.\ee
Thus the bulk term mimics a Wess-Zumino coupling on the brane in
which $B=0$.

It is easy to check explicitly in the case of our example that the
various terms in the tadpole cancellation conditions match up. One
needs the following topological information: $p_1({\mhp}^2) =
2\lambda$, $p_2({\mhp}^2) = 7$, $p_1({\mcp}^2) = 3$,
$p_1(\Lambda^-{\mcp}^2) = -3$. Here $\Lambda^-{\mcp}^2$ denotes
the bundle of anti-self-dual two-forms over ${\mcp}^2$, which is
the normal bundle of ${\mcp}^2$ in $Y={\bf S}^7$. The result for
the first Pontryagin class follows easily since the embedding
space is a seven-sphere. It was also computed explicitly in
Appendix A of \cite{GS}. Substituting these values into the above
formulae, one finds that the gravitational terms are both equal to
$-1/8$. Thus
\be
N_{M2} = -\frac{1}{2}m(m+1)\ee
and \be N_{D2} = -\frac{1}{2}n(n+1)~.\ee
Identifying\footnote{Of course, $N_{M2}$ is allowed to be
negative, corresponding to a non-zero number of space-filling
anti-M2-branes.} $N_{M2}=N_{D2}$, we therefore find that either
$n=m$, or $n=-1-m$ -- the two choices simply correspond to
opposite signs for the flux. Of course, we have shown in a
completely different way that $n=m$, and so we must pick the first
solution. This is therefore an independent check on (\ref{back}).


\section{Freed-Witten Anomaly from K-Theory}

For completeness, in this section we show how one may also derive the shift (\ref{Fshift}) in
the periods of $F$ by
using the K-theory formula \cite{MW, DMW} for the Ramond-Ramond
four-form $G_4$. For simplicity, we set the NS field to zero for the
rest of the paper.

The starting point is the relation
\be
\int_{{\bf S}^2} \frac{G_4}{2\pi} = \frac{F}{2\pi}\label{int}\ee
where recall that ${\bf S}^2$ is any two-sphere linking the
D6-brane worldvolume $Q$. Now, if $G_4$ had periods that were
multiplies of $2\pi$, then one would conclude from (\ref{int})
that $F$ is also standard Dirac quantised. However, Ramond-Ramond
fields are more properly interpreted in K-theory \cite{MW}.
Roughly, $G_4$ -- away from the D6-brane -- is given by the
four-form piece of the Chern character of a K-theory class $x\in
K(Y\setminus Q)$. Here $Y\setminus Q$ denotes $Y$ with the
D6-brane $Q$ deleted. More precisely, we have the following
quantisation condition on $G_4$:
\be \left[\frac{G_4}{2\pi}\right] = \left[\sqrt{\hat{A}(Y\setminus
Q)}\cdot \mathrm{ch}
\left(x+\frac{\Theta}{2}\right)\right]_{\mathrm{four-form}} \in
H^4(Y\setminus Q; \mathbb{Q})\label{K}~.\ee
We will not need to know much about the class $\Theta$. We simply note that,
expanding (\ref{K}), we obtain \cite{DMW}
\be \left[\frac{G_4}{2\pi}\right] = \frac{1}{2}c_1^2(x) - c_2(x) -
\frac{1}{2} \lambda(Y\setminus Q) \quad \mathrm{mod} \
\mz\label{G4}\ee
where $\lambda=p_1/2$. Now, in a tubular neighbourhood of the
D6-brane, this class is a pull-back from $Q$, and so, although at
first sight its contribution in (\ref{G4}) may appear
half-integral, cannot contribute to (\ref{int}). In fact, one can
show\footnote{To see this one needs to note that the fourth Wu
class of $Q$ vanishes on dimensional grounds.} that $\lambda$ is
even in such a tubular neighbourhood. Thus it must be the first
factor in (\ref{G4}) which leads to the half-integer shifts in the
periods of $F$. Indeed, $c_1(x)$ is the first Chern class of the
M-theory circle bundle, and thus may be identified with the
cohomology class of $G_2/2\pi$. Thus we want to compute
\be \int_{{\bf S}^2} \left[\frac{G_2}{2\pi}\right]^2 \quad
\mathrm{mod} \ 2\label{mod2}~.\ee
There are several ways of doing this. One way
is to use an argument similar\footnote{Here the computation was for $[G/2\pi]^2$
over a four-sphere linking an
M5-brane worldvolume.} to the one in section 5.3 of \cite{witten}.
One finds that (\ref{mod2}) is given by $w_2(NQ)$ modulo 2, and therefore
$w_2(Q)$ modulo 2. This is the result we were looking for. However, in this
particular case we can do rather better. Let $SQ$ denote the total
space of the normal
sphere bundle to $Q$ (in other words, the boundary of a tubular neighbourhood, $T$).
 Then one knows the cohomology ring of $SQ$ in terms of that of $Q$ -- it is
given by a polynomial ring
\be
H^*(SQ) \cong H^*(Q)[z]/\left(z^2-c_1(\mathcal{V})z+c_2(\mathcal{V})\right)\ee
where recall that $\mathcal{V}$ is the normal bundle to $Q$ in $X$, viewed as a
complex rank
two vector bundle. Here $z$, which generates the cohomology of the
sphere fibres, may be identified with $c_1(x)$. This formula
follows since $SQ$ is the
projectivisation of $\mathcal{V}$ -- see equation (20.7) of \cite{bott}. Thus we can
compute
\be \int_{{\bf S}^2} z^2 = -\int_{{\bf S}^2}
\left(-c_1(\mathcal{V})z+c_2(\mathcal{V})\right) =
c_1(\mathcal{V})~.\ee
Finally, recall that $c_1(\mathcal{V})$ reduces to $w_2(Q)$,
modulo 2.

\subsection*{Conclusions and some speculative remarks}

By interpreting codimension four fixed point sets in M-theory as
D6-branes in type IIA, and codimension two fixed point sets on membranes
as string boundaries, we have succeeded in deriving the global worldsheet
anomaly for strings ending on a D6-brane, starting from M-theory. Together with
the results of \cite{me}, we have also shown that the Wess-Zumino terms on a
D6-brane may be derived from the Chern-Simons terms in M-theory. It is
amusing to consider\footnote{I would like to thank Sergei Gukov for discussions on
this.} the case of fixed point sets which have a different
(co)dimension. For example, can one make sense of a codimension two fixed
point set? This would naturally become a single-sided boundary in type IIA. Moreover,
strings would appear to be able to end on such an object. It is
tempting to intepret this as some sort of single-sided D8-brane, although
there are many problems with this interpretation. And what about
higher codimension? One runs into an immediate problem for codimension
six, since then taking the projectivisation down to type IIA gives
$\mcp^2$ as fibre. In this case it is not clear how to interpret the fixed point
set in type IIA. Indeed, since $\mcp^2$ does not bound, one even has
problems defining the type IIA manifold.

\centerline{\bf Acknowledgments} I would like to thank J.
Kalkkinen and especially D. Waldram for comments and discussions
on a draft version of this paper. I would also like to thank the
referee of this paper for suggesting that I present more
background on the $G$-index theorem, and also for pointing out
various parts of the text that were not entirely transparent. I am
funded by an EPSRC mathematics fellowship. 

\end{document}